\documentstyle[pre,aps,epsfig,amsfonts,preprint]{revtex} 
\renewcommand{\bbox}[1]{\mbox{\boldmath $#1$}}
\begin{document}                                     
\pagestyle{myheadings}
\markboth{Helbing/Schweitzer/Keltsch/Moln\'{a}r: Active Walker Model for 
Trail Formation}{Helbing/Schweitzer/Keltsch/Moln\'{a}r: 
Active Walker Model for Trail Formation}
\author{Dirk Helbing}
\address{II. Institute of Theoretical Physics, Pfaffenwaldring 57/III, 70550
  Stuttgart, Germany}
\author{Frank Schweitzer}
\address{Institute of Physics, Invalidenstr. 110, 10115 Berlin, Germany}
\author{Joachim Keltsch}
\address{Science+Computing, Hagellocher Weg 71, 72070 T\"ubingen, Germany}
\author{P\'{e}ter Moln\'{a}r}
\address{The Center of Theoretical Studies of Physical Systems,
223 James P. Brawley Drive, 
Atlanta, Georgia 30314, USA}
\title{Active Walker Model for the Formation of Human and 
Animal Trail Systems} 
\maketitle
\begin{abstract}
Active walker models have recently proved their great value for describing
the formation of clusters, periodic patterns, and spiral waves as well as
the development of rivers, dielectric breakdown patterns, and many
other structures. It is shown that they also allow to simulate
the formation of trail systems by pedestrians and ants, 
yielding a better understanding of human and animal behavior.
A comparison with empirical material shows a good agreement between
model and reality.
\par
Our trail formation model includes an equation
of motion, an equation for environmental changes, and an orientation
relation. It contains some model functions, which are specified according to
the characteristics of the considered animals or pedestrians. Not only
the kind of environmental changes differs: Whereas pedestrians leave
footprints on the ground, 
ants produce chemical markings for their orientation.
Nevertheless, it is more important that
pedestrians steer towards a certain destination, while ants usually find their
food sources by chance, i.e. they reach their destination in a stochastic way.
As a consequence, the typical structure of the evolving trail systems depends on the
respective species. Some ant species
produce a dendritic trail system, whereas
pedestrians generate a minimal detour system. 
\par
The trail formation model can be used as a tool for the
optimization of pedestrian facilities: It allows urban planners to design
convenient way systems which actually meet the route choice habits of
pedestrians.
\end{abstract}

\section{Introduction} 

The emergence of complex behavior in a system consisting of simple, interacting
elements \cite{Hak1,Hak2,Pri}
is among the most fascinating phenomena of our world.
Examples can be found in almost every field of today's scientific
interest, ranging from coherent pattern formation in physical and
chemical systems \cite{feisteb-89,kaied-92,claded-95}, to the motion of
animal swarms in biology \cite{pasted-87,deanged-92}, and
the behavior of social groups \cite{valled-94,Weid,Hel3}. 
\par
In the life and social sciences, 
one is usually convinced that the evolution of social
systems is determined by numerous factors, such as cultural,
sociological, economic, political, ecological etc.  However, in recent
years, the development of the interdisciplinary field ``science of
complexity'' has lead to the insight that complex dynamic processes may
also result from simple interactions, and even {\em social} structure formation
could be well described within a mathematical approach
\cite{Weid,Hel3,Ax,Hub,Mont}. 
Moreover, at a certain level of
abstraction, one can find many common features between complex
structures in very different fields. 
\par
A recent field of particular interest is the microsimulation of
self-organization phenomena occuring in traffic systems.
This includes the formation of jammed states in freeway or
city traffic \cite{Nag1,Nag2,Nag3,Nag4,Ban,Kom,Biham,Cuesta,Molera,Naga1,Naga2,Ben,Camp,Hel2}
as well as the various collective patterns of motion developing in
pedestrian crowds \cite{Hel2,BS,Hel1,He,HeMo} like oscillatory changes
of the walking direction at narrow passages or round-about traffic at
crossings.
\par
In this paper, we draw the attention to the specific collective phenomenon
of {\em trail formation} \cite{HeMoSc,Nature}, 
which is widely spread in the world of animals and humans. 
Regarding their shape, duration and extension, trail systems of different
animal species and humans differ, of course. However, more striking is
the question, whether there is a common underlying dynamics which
allows for a generalized description of the formation and evolution of
trail systems. 
\par
As our experience tells us, trails are adapted to the requirements of
their users. In the course of time, frequently used trails become more
developed, making them more attractive, whereas rarely used trails vanish
again. Trails with large detours become optimized by 
creating shortcuts. New destinations or entry points are connected to 
an existing trail system. These dynamical processes occur basically without
any common planning or direct communication 
among the users. Instead, the adaptation
process can be understood as a self-organization phenomenon, resulting 
from the non-linear feedback between
the users and the trails \cite{schw-soweg-96}.
\par
In order to simulate this process, we propose here a 
particle-based, multi-agent
approach to structure formation, which belongs to the class of
{\em active walker models}. Like random walkers, active walkers are subject
to fluctuations and influences of their environment. 
However, they are additionally able to {\em locally change} their environment,
e.g. by altering an environmental potential, which in turn
influences their further movement and their 
behavior. In particular, changes produced by some walkers can influence
other walkers. Hence, the non-linear feedback can be interpreted as an
indirect interaction between the active walkers via environmental
changes, which may lead to the self-organization of spatial structures.
\par
Active walker models have proved their versatility in a variety of
applications, such as formation of complex structures
\cite{Dav,FrLa,Lam1,KaLa,LaPo,Lam2,Schw2}, pattern formation in
physico-chemical systems \cite{Schw,Lutz,Schw1,schw-lsg-mill-96},
aggregation in biological \cite{Stevens,Frank} or urban \cite{schw-steinb-97}
systems, and generation of directed 
motion \cite{eb-schw-ti-ca-95,schw-lao-f-97}.
The approach provides a quite stable and fast
numerical algorithm for simulating processes involving large density
gradients, and it is applicable also in cases where only small particle
numbers govern the structure formation.  In particular, the active walker
model is applicable to processes of pattern formation which are
intrinsically determined by the {\em history} of their creation, such as the
formation of trail systems, discussed in this paper.
\par
In Section II, the active walker model for trail formation is formulated
in terms of a Langevin equation for the movement of the walkers, an
equation for environmental changes, and a relation describing the
orientation of the walkers with respect to 
existing trails. As one application of the model, Section III describes 
the formation of trunk trails in ant colonies, which are commonly used to
exploit food sources. As a second application, in Section IV the
evolution of pedestrian trail systems is modelled. Both Sections III and IV
present a comparison of computational results with real trail systems,
indicating a good aggrement between model and empirical facts. 
In Section IV.A, the equations
for pedestrian trail systems are scaled to dimensionless equations, in order to
demonstrate that the evolving trail systems are (apart from the boundary
conditions) only determined by two parameters.  In Section IV.B, a
macroscopic formulation of human trail formation is derived from the
microscopic equations, allowing analytical investigations and an
efficient calculation of the stationary solution by a self-consistent
field method.  Our conclusions and an outlook, which suggests an application
of the model to the optimization of trail systems, are presented in
Section V.

\section{Active walker model of trail formation}

In order to introduce our model, we first describe the process of trail
formation within a general stochastic framework. Hence, in this
section the active walkers are not specified as pedestrians or animals.
Rather, they are considered as arbitrary moving agents, who 
continuously change their environment by leaving {\em markings} while moving.
These markings can, for example, be imagined as damaged vegetation on
the ground (as in the case of hoofed animals or pedestrians) or as
chemical markings (as in the case of ants).  
\par
The spatio-temporal distribution of the
existing markings will be described by a {\em ground potential}
$G_k(\bbox{r},t)$. Trails are characterized by particularly large
values of $G_k(\bbox{r},t)$. The subscript $k$ allows
to distinguish different {\em kinds} of markings. Due to weathering or 
chemical decay, the markings have a certain {\em life time} $T_k(\bbox{r})$ which 
characterizes their local {\em durability.} Therefore, existing trails tend to fade,
and the ground potential would exponentially adapt to the {\em natural ground 
conditions} $G_k^0(\bbox{r})$, if the production of markings would be stopped.
However, the creation of new markings by agent $\alpha$ 
is described by the term 
$Q_\alpha(\bbox{r}_\alpha,t)\,\delta (\bbox{r} - \bbox{r}_\alpha)$,
where Dirac's delta function $\delta(\bbox{r} -
\bbox{r}_\alpha)$ gives only a contribution at the actual position
$\bbox{r}_\alpha(t)$ of the walker.
The quantity $Q_\alpha(\bbox{r}_\alpha,t)$ represents the strength of new
markings and will be specified later. In summary, we obtain the
following equation for the spatio-temporal evolution of the ground potential:
\begin{eqnarray}
\label{ground1}
 \frac{dG_k(\bbox{r},t)}{dt} &=& \frac{1}{T_k(\bbox{r})}
 \big[ G_k^0(\bbox{r}) - G_k(\bbox{r},t)\big] \nonumber \\
 & & +\; \sum_\alpha Q_\alpha(\bbox{r}_\alpha,t)  
 \, \delta \big(\bbox{r} - \bbox{r}_\alpha(t)\big) \, .
\end{eqnarray}
\par
The {\em motion} of the active walker $\alpha$ on a two-dimensional surface will be
described by the following Langevin equation: 
\begin{mathletters}
\label{langev-a}
\begin{eqnarray}
\frac{d\bbox{r}_{\alpha}(t)}{dt} &=& \bbox{v}_{\alpha}(t) \\ 
\frac{d\bbox{v}_{\alpha}(t)}{dt} & = & -\gamma_\alpha \bbox{v}_{\alpha}(t)
+ \bbox{f}_{\alpha}(t) 
+ \sqrt{2 \,\varepsilon_{\alpha} \, \gamma_\alpha}\, \bbox{\xi}_\alpha(t).
\end{eqnarray}
\end{mathletters}
Eq. (\ref{langev-a}) considers both deterministic and stochastic
influences on the motion of the active walker.  $\bbox{v}_\alpha$ denotes
the {\em actual velocity} of walker $\alpha$. 
$\gamma_\alpha$ represents some kind of
{\em friction coefficient}. 
It is given by the {\em relaxation time} $\tau_\alpha$
of velocity adaptation, specified later: $\gamma_\alpha = 1/\tau_\alpha$. 
The last term describes random variations of the motion in accordance with
the fluctuation-dissipation theorem. $\varepsilon_{\alpha}$ 
is the {\em intensity} of the
stochastic force $\bbox{\xi}_\alpha(t)$, which was assumed to be Gaussian white noise:
\begin{equation}
\langle \bbox{\xi}_\alpha(t) \rangle = \bbox{0} \, , \qquad 
\langle \xi_{\alpha i}(t) \xi_{\beta j}(t') \rangle = 
 \delta_{\alpha\beta} \delta_{ij} \delta (t-t').
\label{noise}
\end{equation}
The $\alpha$-dependence of $\varepsilon_{\alpha}$ takes into account 
that different walkers could behave more or less erratic,
dependent on their current situation. 
\par
Finally, the term $\bbox{f}_{\alpha}$ represents deterministic 
influences on the motion, such
as intentions to move into a certain direction with a certain desired
velocity, or to keep distance from
neighboring walkers. According to the {\em social force concept}
\cite{Hel2,Hel1}, $\bbox{f}_{\alpha}$ is specified as follows:
\begin{eqnarray}
 \bbox{f}_{\alpha}(t) 
&=& \frac{v_\alpha^0}{\tau_\alpha} 
\bbox{e}_\alpha(\bbox{r}_\alpha,\bbox{v}_\alpha,t)
\nonumber \\
& & + \sum_{\beta(\ne \alpha)} \bbox{f}_{\alpha\beta}(\bbox{r}_\alpha,
 \bbox{v}_\alpha;\bbox{r}_\beta,\bbox{v}_\beta) \, .
\label{fortbew}
\end{eqnarray}
Here, $v_\alpha^0$ describes the {\em desired velocity} and
$\bbox{e}_\alpha$ the {\em desired direction} of the walker. The term
$\bbox{f}_{\alpha\beta}$ delineates the effect of pair interactions between
walkers $\alpha$ and $\beta$ on the motion of walker $\alpha$
\cite{Hel2,Hel1,HeMoSc}. Since we will focus on cases of rare direct
interactions, $\bbox{f}_{\alpha\beta}$ can be approximately neglected
here. Thus, eq. (\ref{langev-a}b) becomes 
\begin{equation}
\frac{d\bbox{v}_{\alpha}(t)}{dt} = \frac{v_\alpha^0
  \bbox{e}_\alpha(\bbox{r}_\alpha,\bbox{v}_\alpha,t) - \bbox{v}_\alpha(t)}{\tau_\alpha}
 + \sqrt{2 \,\varepsilon_{\alpha} \, \gamma_\alpha}\, \bbox{\xi}_\alpha(t) \, ,
\label{langev-b}
\end{equation}
where the first term reflects an adaptation of the actual walking
direction $\bbox{v}_\alpha / \|\bbox{v}_\alpha \|$
to the desired walking direction $\bbox{e}_\alpha$ and an
acceleration towards the desired velocity $v_\alpha^0$ with a certain
relaxation time $\tau_\alpha$. Assuming that the time $\tau_\alpha$ is rather short
compared to the time scale of trail formation (which is characterized by the
durability $T_k$), equation 
(\ref{langev-b}) can be adiabatically eliminated. This leads to the following
{\em equation of motion:}
\begin{equation}
\label{reduct}
  \frac{d\bbox{r}_\alpha}{dt} = \bbox{v}_\alpha(\bbox{r}_\alpha,t)
  \approx v_\alpha^0 \bbox{e}_\alpha(\bbox{r}_\alpha,\bbox{v}_\alpha,t) 
  + \sqrt{2 \,\varepsilon_{\alpha} \tau_\alpha}\, \bbox{\xi}_\alpha(t) \, .
\end{equation}
\par
To complete our trail formation model, we must finally specify the
{\em orientation relation}
\begin{equation}
 \bbox{e}_\alpha(\bbox{r}_\alpha,\bbox{v}_\alpha,t) = \bbox{e}_\alpha(
 \{G_k(\bbox{r},t)\},\bbox{r}_\alpha,\bbox{v}_\alpha) \, ,
\end{equation}
which determines the desired walking direction in dependence of the
ground potentials $G_k(\bbox{r},t)$. Since the concrete 
orientation relation for pedestrians differs from that
for ants, it will be introduced later on, in the respective sections.
However, it is clear that the presence of a trail will have an
{\em attractive} effect, i.e. it will induce an orientation {\em towards} it.
According to Eq. (\ref{reduct}), this will cause a tendency to approach and
to use the trail.
\par
Therefore, the mechanism of trail formation 
is based on some kind of {\em agglomeration
process}, which is {\em delocalized} due to the directedness of the walkers' motion.
Starting with a plain, spatially homogeneous ground, the walkers will
move arbitrarily. However, by continuously leaving markings, they produce trails
which have an attractive effect on nearby walkers. Thus, the agents
begin to use already existing trails after some time.
By this, a kind of {\em selection process} between trails 
occurs (cf.~\cite{Schw}): Frequently used trails are reinforced, which makes
them even more attractive, whereas rarely used trails may vanish 
again. The trails begin to
bundle, especially where different trails meet or intersect. Therefore,
even walkers with different entry points
and destinations use and maintain common parts of the trail system.

\section{Trunk trail formation by ants}

As a first example, we want to model the formation of trunk trails,
which is a widely observed phenomenon in ant colonies, such as in the
{\em Myrmicinae, Dolichoderinae} and {\em Formicinae} species, commonly
foraging for food from a central nest
\cite{schw-lao-f-97,Hoell,hoelld-wilson-90}. The trails are used to
connect the food sources with the nest to allow for a collective
exploitation of the food. In the case of ants, the markings 
are chemical signposts, so-called {\em pheromones}, which also provide
the basic orientation for foraging and homing of the animals.  However, note
that not all ants species form trails. There is a variety of
very complex foraging patterns in ants, such as swarm riding of army ants
(e.g. in the species of {\em Eciton} and {\em Dorylus}) \cite{deneub}.
Therefore, we restrict here to cases, in which trunk trail
formation of group riding ants is reported.
\par
Before we present our model, we would like to mention some differences between
active walkers and ants. The latter are rather complex biological
creatures which are capable of using additional information (e.g.
landmark use) or egocentric navigation \cite{wehner-92} for their food
searching and homing. Moreover, they can store information in an
individual memory and communicate with nest mates in a very complex
manner \cite{haefner-crist-94}. 
\par
We will neglect these abilities,
in order to show that they are not necessary for trail formation.
The active walkers in our
model merely count on the {\em local information} provided by the
chemical trail, in order to guide themselves. They do not have additional
navigation or information processing capabilities, and are not subject to
long-range attracting forces to the food sources or to the nest. Hence,
the formation of trunk trails in the following model is clearly a self-organizing
process, based on the local interactions of the walkers
\cite{schw-lao-f-97}. 
\par
Trunk trails used for foraging are typically dendritic in form. Each one
starts from the nest vicinity as a single thick pathway that splits first
into branches and then into twigs to convey  
large numbers of ants rapidly
into the foraging areas (see Fig.~1). 
\par
In order to
distinguish those trails which lead to a food source, the ants, {\em
  after} discovering a food source, use {\em another} pheromone to mark
their trails, which stimulates the recruitment of additional ants to
follow that trail. In our active walker model, we count on that fact by
using two different chemical markings: Chemical $0$ is used by the
active walkers as long as they have not reached a food source, i.e. on
their way from the nest to the food or during search periods. Chemical
$1$ is only used by active walkers after they have reached a food source,
i.e. on their way back from food sources to the nest. An internal
parameter $k_\alpha=\{0,1\}$ 
indicates which of these markings is produced by the active walker $\alpha$. 
Hence, the production term for the ground potential in 
Eq. (\ref{ground1}) is defined as follows:
\begin{eqnarray}
Q_{\alpha}(\bbox{r}_\alpha,t) &=& 
 (1-k_\alpha) q_0 \exp [-\beta_0\,(t-t^\alpha_0)] \nonumber \\
& & \qquad + k_\alpha q_1 \exp [-\beta_1\,(t-t^\alpha_1)] \, .
\label{prod}
\end{eqnarray}
The first term is relevant for $k_\alpha = 0$, i.e. when searching a food
source, whereas the second term contributes only for $k_\alpha = 1$,
i.e. after having found some food. Since the capacity of producing 
chemical markings is limited, we have assumed that
the quantity of chemical produced by a
walker after leaving the nest or the food source decreases
exponentially in time, where $\beta_0$ and $\beta_1$ 
are the respective decay parameters.
$q_0$ and $q_1$ denote the initial production, and 
$t^\alpha_0$, $t^\alpha_1$ are the times, when the
walker $\alpha$ has started from the nest or the food source, respectively.
\par
Due to the two chemical markings, we have two different ground potentials 
$G_0(\bbox{r},t)$ and $G_1(\bbox{r},t)$ here, which provide orientation for 
the walkers. In the following, we need to specify how they influence the
motion of the agents,
especially their desired directions $\bbox{e}_\alpha(\bbox{r}_\alpha,
 \bbox{v}_\alpha,t)$. 
At this point, we take into account that 
the walkers $\alpha$ will not directly be affected in their behavior by the
ground potentials $G_k(\bbox{r},t)$ themselves, which 
reflect the pure {\em existence}
of markings of type $k$ at place $\bbox{r}$. They will rather be influenced
by the {\em perception} of their environment from their actual positions
$\bbox{r}_\alpha$, which will be described by the {\em trail potentials}
\begin{equation}
 V_{tr}^k(\bbox{r}_\alpha,\bbox{v}_\alpha,t)
 = V_{tr}^k(\{G_k(\bbox{r},t)\},\bbox{r}_\alpha,\bbox{v}_\alpha) \, .
\end{equation}
\par
For the detection of chemical markings, insects like ants use specific
receptors which are located at their so-called antennae.
Their perception is mainly determined by the angle $2\varphi$ of perception, 
which is given by the angle between the antennae (cf. Fig.~2). 
Therefore, we make the assumption
\begin{eqnarray}
\label{trail-ant}
 & & \hspace*{-3mm} V_{tr}^k(\bbox{r}_\alpha,\bbox{v}_\alpha,t) \nonumber \\
 & & = \int_0^{\Delta r} \!\!dr'\int_{-\varphi}^{+\varphi} \!\!d \varphi'\, 
 r' \nonumber \\
& & \times \; G_k\Big(\bbox{r}_{\alpha}+ r' 
\big(\cos(\omega_\alpha+\varphi'),\sin(\omega_\alpha+\varphi')\big),t\Big) 
 \, ,
\end{eqnarray}
where the angle $\omega_\alpha$ is given by the current walking direction
\begin{equation}
  \bbox{e}_\alpha^*(t) = \frac{\bbox{v}_\alpha(t)}{\|\bbox{v}_\alpha(t) \|}
 = \big( 
 \cos \omega_\alpha(t) , 
 \sin \omega_\alpha(t) \big) \, .
\end{equation}
According to (\ref{trail-ant}), our active walkers integrate over the ground 
potential between the antennae of length $\Delta r$. Note, however, 
that the restriction to the
angle of perception is not an indispensible assumption for the generation
of trails \cite{Tilch}. Thus, it could be neglected in a minimal model.
Nevertheless, it has been introduced to mimic the biological constitution 
of the ants and to keep close to biology. 
\par
The perception of already existing trails will have an {\em attractive
  effect} $\bbox{f}_{tr}(\bbox{r}_\alpha,\bbox{v}_\alpha,t)$ to the active
walkers. This has been defined by the gradients of the trail potentials, 
\begin{eqnarray}
\bbox{f}_{tr}^\alpha(\bbox{r},\bbox{v},t) 
 &=& (1- k_\alpha) \bbox{\nabla} V_{\rm
   tr}^1(\bbox{r},\bbox{v},t) \nonumber \\
 & & \qquad + k_\alpha \bbox{\nabla} V_{tr}^0 (\bbox{r},\bbox{v},t) \, .
\end{eqnarray}
The above formula takes into account that 
walkers which move out from the nest to reach a food source ($k_\alpha=0$) 
orientate by chemical $1$,
whereas walkers which move back from the food ($k_\alpha=1$)
orientate by chemical $0$. This implies that initially, in the absense of
chemical $1$, the walkers move as random walkers which discover a food
source only by chance. 
\par
We complete our model of trunk trail formation by specifying the
{\em orientation relation} of the walkers. Assuming 
$\bbox{e}_\alpha(\bbox{r},\bbox{v},t)
 = \bbox{f}_{tr}^\alpha(\bbox{r},\bbox{v},t)
 / \| \bbox{f}_{tr}^\alpha(\bbox{r},\bbox{v},t) \|$,
the desired walking direction $\bbox{e}_\alpha(\bbox{r},\bbox{v},t)$ 
points into the direction of the steepest increase of the relevant
trail potential $V_{tr}^k(\bbox{r},\bbox{v},t)$. However, this formula
does not take into account the ants' persistence to keep the previous 
direction of motion \cite{Alt}. The latter reduces the probability
of changing to the opposite walking direction by fluctuations, 
which would cause the ants to move backwards before reaching their goal.
Therefore, we modify the above formula to
\begin{equation}
\label{Orient2}
 \bbox{e}_\alpha(\bbox{r},\bbox{v},t)
 = \frac{\bbox{f}_{tr}^\alpha(\bbox{r},\bbox{v},t)
 + \bbox{e}_\alpha^*(t-\Delta t) }{{\cal N}_\alpha (\bbox{r},\bbox{v},t)}
 \, ,
\end{equation}
where ${\cal N}_\alpha(\bbox{r},\bbox{v},t) 
= \| \bbox{f}_{tr}^\alpha(\bbox{r},\bbox{v},t) 
 + \bbox{e}_\alpha^*(t-\Delta t) \|$ is a normalization factor.
That means, on a ground without markings, the
walking direction tends to agree with the one at the previous
time $t-\Delta t$, but
it can change by fluctuations.
\par
Finally, it is known from ant species that they are able to leave a place where
they do not find food and increase their mobility to reach out for other
areas. Since active walkers do not reflect their situation, they stick on
their local markings even if they did not find any food source.  In
order to increase the mobility of the active walkers in those cases, we
assume that every walker has an individual noise intensity
$\varepsilon_{\alpha}$(t), which is related
to the walker's spatial diffusion coefficient and should increase
continuously, as long as the walker does not find a food source:
\begin{eqnarray}
\varepsilon_\alpha(t)&=&
(1 - k_\alpha)[\varepsilon_0+r_\varepsilon (t-t^\alpha_0)]^{2}  
\nonumber \\
& &\qquad + k_\alpha \varepsilon_{0}^{\;2} \, .
\label{si}
\end{eqnarray}
$t^\alpha_0$ is again the starting time of walker $\alpha$ from the nest,
$\varepsilon_0$ is the initial noise level and $r_\varepsilon$ its growth rate.
If the noise intensity $\varepsilon_{\alpha}$ has reached a critical upper
value, the walker $\alpha$ 
behaves more or less as a random walker which does not pay attention to the trail
potential. But if the walker found some food, 
its individual noise intensity is set back to the initial value $\varepsilon_0$.
\par
Figure~1b shows the result of computer simulations of trunk trail
formation. The related dendritic trail system of {\em Pheidole
  milicida}, a harvesting ant of the southwestern U.S. deserts, is displayed
in Figure~1a. In our
simulation, a nest is assumed in the middle of a triangular lattice of
size $100 \times 100$ with periodic boundary conditions. Initially, there
are no chemical markings on the lattice. At time $t=0$, a number
$N_0$ of walkers start from the nest with a random direction, leaving
markings of chemical $0$. If a walker disovers a food source by chance,
it begins to produce chemical $1$. Should such a walker find its way back
to the nest, it activates an additional number of walkers, the recruits, to
move out. The maximum number of walkers in the simulation is limited to
$N_{max}$, which denotes the population size.
\par
For the food sources, an extended food distribution at the top and bottom
line of the lattice is assumed \cite{Note}.
These sources could be exhausted by the
visiting walkers, but the accidental discovery of new ones in the neighborhood
results in a branching of the main trails in the vicinity of the food
sources and eventually leads to the dendritic structures. The trail
system observed in Figure~1b  remains unchanged in its major
parts as has been reported also in
the biological observations of trunk trail formation by ants \cite{Hoell}.
Nevertheless, some minor trails in the vicinity of the food sources 
slightly shift in the course of time due to fluctuations.

\section{Human trail formation}

Trail formation by pedestrians has been investigated only very 
recently \cite{Schenk}.
It can be interpreted as a complex interplay
between pedestrian motion, human orientation, and environmental changes:
On the one hand, pedestrians tend to take the shortest way to their
destination. On the other hand, they avoid to walk on bumpy ground, since
this is uncomfortable. Therefore, they prefer to use existing trails, but
they build a new shortcut, if the relative detour would be too large.  In
the latter case they generate a new trail, since footprints
clear some vegetation. Examples of the resulting trail systems can be
found in green areas, like public parks (cf.  Fig.~3).
\par
Empirical studies have shown that pedestrian motion can be
surprisingly well described by the {\em social force
  model} sketched in Section~II \cite{Hel3,BS}. 
In particular, it has been demonstrated
that this model allows a realistic simulation of various observed
self-organization phenomena in pedestrian crowds
\cite{Hel2,BS,Hel1,He,HeMo,HeMoSc}. This includes the emergence of
collective patterns of motion, e.g. lanes of uniform walking direction
\cite{Hel1,HeMoSc} or roundabout traffic at intersections
\cite{He,HeMo,HeMoSc}. 
\par
In this section, however, we want to model the evolution of human trail
patterns. We will assume that the pedestrians behave `reasonably' and, as
before, we will restrict our model to the most important factors.
It is obvious that pedestrians are able to show a much more complicated 
behavior than described here.
\par 
Since the equation (\ref{reduct}) of motion can be also applied to pedestrians,
we have to specify how moving pedestrians
change their environment by leaving footprints, now. This time, we do not have to
distinguish different kinds of markings. Thus, we will need only 
one ground potential $G(\bbox{r},t)$, and the subscript $k$ can
be omitted. The value of $G$ is 
a measure of the {\em comfort of walking}.
(Therefore, it can considerably depend on the weather conditions, which is
not discussed here any further.)
\par
For the strength $Q_\alpha(\bbox{r},t)$ of the
markings produced by footprints at place $\bbox{r}$ we assume
\begin{equation}
  \label{q-ped}
Q_\alpha(\bbox{r},t)=I(\bbox{r})\; \left[ 1 - \frac{G(\bbox{r},t)}{G_{max}(\bbox{r})}
\right]\, ,   
\end{equation}
where $I(\bbox{r})$ is the location-dependent intensity of clearing vegetation. 
The saturation term $[1 - G(\bbox{r},t)/ G_{max}(\bbox{r})]$ results from the
fact that the clarity of a trail is limited to a maximum value
$G_{max}(\bbox{r})$.
\par
On a plain, homogeneous ground without any trails, the desired
direction $\bbox{e}_\alpha$ of a pedestrian $\alpha$ at place
$\bbox{r}$ is given by the direction $\bbox{e}_\alpha^*$
of the next destination $\bbox{d}_\alpha$, i.e.
\begin{equation}
 \bbox{e}_\alpha(\bbox{r},\bbox{v},t) =
 \bbox{e}_\alpha^*(\bbox{d}_\alpha,\bbox{r}) 
 =\frac{\bbox{d}_\alpha - \bbox{r}}{\| \bbox{d}_\alpha - \bbox{r} \|}
 = \bbox{\nabla} U_\alpha(\bbox{r})
\end{equation}
with the {\em destination potential}
\begin{equation}
 U_\alpha(\bbox{r}) = - \| \bbox{d}_\alpha - \bbox{r} \| \, .
\end{equation}
\par
However, the perception of already existing trails will have an {\em attractive
  effect} $\bbox{f}_{tr}(\bbox{r},t)$ on the walker,
which will again be defined by the gradient of the trail potential
$V_{tr}(\bbox{r},t)$, specified later on:
\begin{equation}
\bbox{f}_{tr}(\bbox{r},t) = \bbox{\nabla} V_{tr}
 (\bbox{r},t) \, .
\end{equation}
Since the potentials $U$ and $V_{tr}$ influence the pedestrian at the
same time, it seems reasonable to introduce an 
{\em orientation relation} similar to
(\ref{Orient2}), by taking the sum of both potentials: 
\begin{eqnarray}
\label{ea}
 \bbox{e}_\alpha(\bbox{r},\bbox{v},t) 
 &=& \frac{\bbox{f}_{tr}(\bbox{r},t) + \bbox{e}_\alpha^*(\bbox{d}_\alpha,
 \bbox{r})}{{\cal N}(\bbox{r},t)} \nonumber \\
 &=& \frac{1}{{\cal N}(\bbox{r},t)} 
\bbox{\nabla} [ U_\alpha(\bbox{r}) + V_{tr}(\bbox{r},t) ]
\end{eqnarray}
Here, ${\cal N}(\bbox{r},t) = \| \bbox{\nabla} [
U_\alpha(\bbox{r}) + V_{tr}(\bbox{r},t) ] \|$ serves as
normalization factor. By relation (\ref{ea}) we reach that 
the vector $\bbox{e}_\alpha(\bbox{r}_\alpha,t)$ points into
a direction which is 
a compromise between the {\em shortness} of the 
direct way to the destination 
and the {\em comfort} of using an existing trail. 
\par
Finally, we need to specify the trail potential $V_{tr}$ for pedestrians.
Obviously a trail must be recognized by the walkers and near enough in order to
be used. Whereas the ground potential $G(\bbox{r},t)$ describes the {\em
  existence} of a trail segment at position $\bbox{r}$, the {\em trail potential}
$V_{tr}(\bbox{r}_\alpha,t)$ reflects the {\em attractiveness} of a trail
from the actual position $\bbox{r}_\alpha(t)$ of the walker. Since this will
decrease with the distance $\|\bbox{r} - \bbox{r}_\alpha \|$, we have 
applied the relation
\begin{equation}
\label{trail-ped}
 V_{tr}(\bbox{r}_\alpha,t) = \int d^2 r\, \mbox{e}^{-\|\bbox{r} - 
 \bbox{r}_\alpha\|/\sigma(\bbox{r}_\alpha)} G(\bbox{r},t) \, ,
\end{equation}
where $\sigma(\bbox{r}_\alpha)$ characterizes the sight, i.e. the range 
of visibility. 
In analogy to (\ref{trail-ant}), this formula could be easily 
generalized to include conceivable effects of a pedestrian's
angle of sight. However, we will not do this here, since we would have to calculate
different trail potentials $V_{tr}^\alpha$ for all walkers $\alpha$, then.
This would make the model much more complicated.
\par
The simulation results of the above described trail formation model
are in good agreement with empirical observations, as can be seen
by comparison with photographs. Our
{\em multi agent simulations} begin with plain, homogeneous ground. All
pedestrians have their own destinations and entry points (like shops,
houses, underground stations, or parking lots), from which they start at
a randomly chosen time.  In Figure~4 the entry points
and destinations are distributed over the small ends of the ground,
while in Figure~5 (Fig.~7) pedestrians move
between all possible pairs of three (four) fixed places.
\par
At the beginning, pedestrians take the {\em direct ways} to their
respective destinations. However, after some time pedestrians begin to
use already existing trails, since this is more comfortable than to clear
new ways. The frequency of usage decides which trails are reinforced
and which ones vanish in the course of time.
If the attractiveness of the forming trails is large, the final trail
system is a {\em minimal way system} (which is the shortest way system
that connects all entry points and destinations). However, because of the
pedestrians' dislike of taking detours the evolution of the trail system
normally stops before this state is reached. In other words, a so-called
{\em minimal detour system} develops if the model parameters are chosen
realistically (cf. Fig.~5). The resulting trails can considerably
differ from the direct ways which the pedestrians would use if these
were equally comfortable.

\subsection{Scaling to dimensionless equations}

The use of existing trails depends on the visibility, as given by Eq.
(\ref{trail-ped}). Assuming that the sight parameter $\sigma$ is
approximately space-independent, an additional simplification of the
equations of trail formation can be reached by introducing dimensionless
variables
\begin{eqnarray}
 \bbox{x} &=& \frac{\bbox{r}}{\sigma} \, , \\
 \tau(\bbox{x}) &=& \frac{t}{T(\sigma\bbox{x})} \, , \\
 G'(\bbox{x},\tau) &=& \sigma G(\sigma \bbox{x},\tau T)
 \, , \\
 V'_{tr}(\bbox{x},\tau) &=& \int d^2 x' \, \mbox{e}^{-\|\bbox{x}' - \bbox{x} \|}
 G'(\bbox{x}',\tau) \, , \label{V_tr} \\
 U'_\alpha(\bbox{x}) &=& - \| \bbox{d}_\alpha /\sigma - \bbox{x}
 \| \, ,
\end{eqnarray}
etc.
Neglecting fluctuations in equation (\ref{reduct}) for the moment,
this implies the following scaled equations:
\begin{equation}
 \frac{d\bbox{x}_\alpha(\tau)}{d\tau} 
 = \bbox{v}_\alpha^{\,\prime}(\bbox{x}_\alpha,\tau) 
 \approx \frac{v_\alpha^0 T(\sigma\bbox{x}_\alpha)}
 {\sigma} \bbox{e}_\alpha^{\,\prime}(\bbox{x}_\alpha,\tau) 
\end{equation}
for pedestrian motion,
\begin{equation}
 \bbox{e}_\alpha^{\,\prime}(\bbox{x},\tau) = 
\frac{\bbox{\nabla} [ U'_\alpha(\bbox{x})
 + V'_{tr}(\bbox{x},\tau) ]} 
 {\| \bbox{\nabla} [ U'_\alpha(\bbox{x})
 + V'_{tr}(\bbox{x},\tau) ] \|}
\end{equation}
for human orientation, and
\begin{eqnarray}
 \frac{dG'(\bbox{x},\tau)}{d\tau}
 &=& \big[ G'_0(\bbox{x}) - G'(\bbox{x},\tau)\big] + \left[ 1 - 
 \frac{G'(\bbox{x},\tau)}{G'_{max}(\bbox{x})} \right] \nonumber \\
 & & \times \sum_\alpha \frac{I(\sigma\bbox{x}) T(\sigma\bbox{x})}{\sigma} 
 \, \delta \big(\bbox{x} - \bbox{x}_\alpha(\tau)\big) 
\label{envch}
\end{eqnarray}
for environmental changes.
Therefore, we find the surprising result that the dynamics of
trail formation is (apart from the influence of the number and
places of entering and leaving pedestrians)
already determined by two local parameters $\kappa$
and $\lambda$ instead of four, namely the products
\begin{equation}
 \kappa(\bbox{x}) = \frac{I(\sigma \bbox{x}) T(\sigma \bbox{x})}{\sigma} 
\end{equation}
and
\begin{equation}
 \lambda(\bbox{x}) = \frac{V^0T(\sigma\bbox{x})}{\sigma} \, .
\end{equation}
Herein, $V^0$ denotes the mean value of the desired velocities
$v_\alpha^0$.

\subsection{Macroscopic formulation of trail formation}

 From the above `microscopic' model of trail formation we will now derive
the related `macroscopic' equations. For this purpose we need to
distinguish different subpopulations $a$ of individuals $\alpha$. By
$a(\tau)$ we denote the time-dependent set of individuals $\alpha$ who
have started from the same entry point $\bbox{p}_a$ with the same
destination $\bbox{d}_a$.  Therefore, the different sets $a$ correspond to
the possible (directed) combinations between existing entry points and
destinations.
\par
Next, we define the {\em density} $\rho_a(\bbox{x},\tau)$ of individuals
of subpopulation $a$ at place $\bbox{x}$ by
\begin{equation}
 \rho_a(\bbox{x},\tau) = 
 \sum_{\alpha \in a(\tau)}
 \delta \big(\bbox{x} - \bbox{x}_\alpha(\tau)\big) \, .
\label{defin}
\end{equation}
\par
Note that a spatial smoothing of the density is reached by a
discretization of space, which is needed for a numerical implementation
of the model.  For example, if the discrete places $\bbox{x}_i$ represent
quadratic domains
\begin{equation}
 {\cal A}(\bbox{x}_i) = \{ \bbox{x} : \| \bbox{x} - \bbox{x}_i \|_\infty \le L \}
\end{equation}
with an area $|{\cal A}| = L^2$, the corresponding density is 
\begin{equation}
  \rho_a(\bbox{x}_i,\tau) = 
 \frac{1}{|{\cal A}|} \int\limits_{{\cal A}(\bbox{x}_i)} d^2 x 
 \sum_{\alpha \in a(\tau)} \delta \big(\bbox{x} - \bbox{x}_\alpha(\tau)\big) \, .
\end{equation}
However, for reasons of simplicity we will treat the continuous case.
\par
The quantity
\begin{eqnarray}
 N_a(\tau) &=& \int d^2 x \, \rho_a(\bbox{x},\tau) \nonumber \\
 &=& \sum_{\alpha \in a(\tau)} \int d^2 x \,  
 \delta \big(\bbox{x} - \bbox{x}_\alpha(\tau)\big) 
\end{eqnarray}
describes the number of pedestrians of subpopulation $a$, who are walking
on the ground at time $\tau$. It changes by pedestrians entering the
system at the entry point $\bbox{p}_a$ with a rate $R_a^+(\bbox{p}_a,\tau)$
and leaving it at the destinations $\bbox{d}_a$ with a rate
$R_a^-(\bbox{d}_a,\tau)$.
\par
Due to the time-dependence of the sets $a(\tau)$, we will need 
the set 
\begin{equation}
 a_\cap(\tau) = a(\tau + \Delta) \cap a(\tau) 
\end{equation}
of pedestrians remaining in the system, the set
\begin{equation}
 a_+(\tau) = a(\tau + \Delta) \setminus a_\cap(\tau) 
\end{equation}
of entering pedestrians, and the set
\begin{equation}
 a_-(\tau) = a(\tau) \setminus a_\cap(\tau) 
\end{equation}
of leaving pedestrians,
for which the following relations hold:
\begin{eqnarray}
 a_+(\tau) \cap a_-(\tau) &=& \emptyset \, , \\
 a_\cap(\tau) \cup a_+(\tau) &=& a(\tau + \Delta) \, , \\
 a_\cap(\tau) \cup a_-(\tau) &=& a(\tau) \, .
\end{eqnarray}
Therefore, equation (\ref{defin}) implies 
\begin{eqnarray}
 \frac{\partial\rho_a(\bbox{x},\tau)}{\partial \tau} 
  &=& \lim_{\Delta \rightarrow 0} \frac{1}{\Delta}
 \left[ \sum_{\alpha \in a(\tau+\Delta)} \delta\big(\bbox{x} - \bbox{x}_\alpha(\tau
   + \Delta)\big) \right. \nonumber \\
 & & \qquad \quad \left. - \sum_{\alpha \in a(\tau)} \delta\big(\bbox{x} -
   \bbox{x}_\alpha(\tau)\big) \right] \nonumber \\
 &=& \lim_{\Delta \rightarrow 0} \sum_{\alpha \in a_\cap(\tau)} \frac{1}{\Delta}
 \Big[ \delta\big(\bbox{x} - \bbox{x}_\alpha(\tau+\Delta)\big) \nonumber \\
 & & \qquad \quad - \delta\big(\bbox{x} - \bbox{x}_\alpha(\tau)\big) \Big] \nonumber \\
 & & + \lim_{\Delta \rightarrow 0} \frac{1}{\Delta} 
 \sum_{\alpha \in a_+(\tau)} \delta\big(\bbox{x} - \bbox{x}_\alpha(\tau + 
 \Delta)\big) 
 \nonumber \\
 & & - \lim_{\Delta \rightarrow 0} \frac{1}{\Delta} 
 \sum_{\alpha \in a_-(\tau)} \delta\big(\bbox{x} - \bbox{x}_\alpha(\tau)\big) 
 \, .
\end{eqnarray}
Taking into account
\begin{eqnarray}
 \lim_{\Delta \rightarrow 0} \frac{1}{\Delta}
 \delta\big(\bbox{x} - \bbox{x}_\alpha(\tau + \Delta)\big)
 &=& \lim_{\Delta \rightarrow 0} \frac{1}{\Delta}
 \delta\big(\bbox{x} - \bbox{x}_\alpha(\tau)\big) \nonumber \\
 & & + \frac{\partial}{\partial \tau} 
 \delta\big(\bbox{x} - \bbox{x}_\alpha(\tau)\big) \, ,
\end{eqnarray}
which follows by Taylor expansion, we obtain
\begin{eqnarray}
 \frac{\partial\rho_a(\bbox{x},\tau)}{\partial \tau} 
 &=& \lim_{\Delta \rightarrow 0} \sum_{\alpha \in a(\tau+\Delta)} 
\frac{1}{\Delta}
 \Big[ \delta\big(\bbox{x} - \bbox{x}_\alpha(\tau+\Delta)\big) \nonumber \\
 & & \qquad \quad - \delta\big(\bbox{x} - \bbox{x}_\alpha(\tau)\big) \Big] \nonumber \\
 & & + \lim_{\Delta \rightarrow 0} \frac{1}{\Delta} 
 \sum_{\alpha \in a_+(\tau)} \delta\big(\bbox{x} - \bbox{x}_\alpha(\tau)\big)
 \nonumber \\
 & & - \lim_{\Delta \rightarrow 0} \frac{1}{\Delta} 
 \sum_{\alpha \in a_-(\tau)} \delta\big(\bbox{x} - \bbox{x}_\alpha(\tau)\big) 
\, .
\end{eqnarray}
With
\begin{eqnarray}
 & & \hspace*{-0.5cm} \lim_{\Delta \rightarrow 0} \frac{1}{\Delta}
 \Big[ \delta\big(\bbox{x} - \bbox{x}_\alpha(\tau+\Delta)\big) 
 - \delta\big(\bbox{x} - \bbox{x}_\alpha(\tau)\big) \Big] \nonumber \\
 &=& \frac{\partial}{\partial \tau} \delta\big(\bbox{x} - \bbox{x}_\alpha(\tau)\big)
 \nonumber \\
 &=& - \bbox{\nabla} \delta\big(\bbox{x} - \bbox{x}_\alpha(\tau)\big)
 \cdot \frac{d\bbox{x}_\alpha}{d\tau}
\end{eqnarray}
and the relations
\begin{eqnarray}
 R_a^+(\bbox{x},\tau) &=& 
 \lim_{\Delta \rightarrow 0} \frac{1}{\Delta} \sum_{\alpha \in a_+(\tau)} 
 \delta\big(\bbox{x} - \bbox{x}_\alpha(\tau)\big)
 \\
 R_a^-(\bbox{x},\tau) &=& 
 \lim_{\Delta \rightarrow 0} \frac{1}{\Delta} \sum_{\alpha \in a_-(\tau)} 
 \delta\big(\bbox{x} - \bbox{x}_\alpha(\tau)\big) 
\end{eqnarray}
for the rates of pedestrians joining and leaving subpopulation $a$,
we finally arrive at
\begin{eqnarray}
 \frac{\partial\rho_a(\bbox{x},\tau)}{\partial \tau} 
 &=& - \bbox{\nabla} \cdot 
 \sum_{\alpha \in a(\tau)} \bbox{v}_\alpha^{\,\prime}(\bbox{x}_\alpha,\tau)
 \, \delta\big(\bbox{x} - \bbox{x}_\alpha(\tau)\big) \nonumber \\
 & & + R_a^+(\bbox{x},\tau) - R_a^-(\bbox{x},\tau) \, ,
\end{eqnarray}
where $R_a^+(\bbox{x},\tau)$ is zero away from the entry point $\bbox{p}_a$,
and the same holds for $R_a^-(\bbox{x},\tau)$ 
away from the destination $\bbox{d}_a$.
\par
Now, we define the {\em average velocity} $\bbox{V}_a$ by 
\begin{equation}
 \bbox{V}_a(\bbox{x},\tau) = \frac{1}{\rho_a(\bbox{x},\tau)} 
 \sum_{\alpha \in a(\tau)} \bbox{v}_\alpha^{\,\prime}(\bbox{x}_\alpha,\tau)
 \delta\big(\bbox{x} - \bbox{x}_\alpha(\tau)\big) \, .
\end{equation}
This gives us the desired {\em continuity equation}
\begin{eqnarray}
 \frac{\partial\rho_a(\bbox{x},\tau)}{\partial \tau} 
 &+& \bbox{\nabla} \cdot 
 \big[ \rho_a(\bbox{x},\tau) \bbox{V}_a(\bbox{x},\tau) \big] \nonumber \\
 & & = R_a^+(\bbox{x},\tau) - R_a^-(\bbox{x},\tau) 
\label{cont}
\end{eqnarray}
describing pedestrian motion. Fluctuation effects can be taken into
account by the additional {\em diffusion terms}
\begin{equation}
 \sum_b \bbox{\nabla} \cdot \big[ D_{ab}(\{\rho_c\}) \bbox{\nabla}
 \rho_b(\bbox{x},\tau) \big]
\end{equation}
on the right-hand side of (\ref{cont}) \cite{Grein,Hel2}. This causes the
trails to become somewhat broader.
\par
Next, we rewrite equation (\ref{envch}) for environmental changes in the
form
\begin{eqnarray}
 \frac{dG'(\bbox{x},\tau)}{d\tau}
 &=& \big[ G'_0(\bbox{x}) - G'(\bbox{x},\tau)\big] \nonumber \\
 & & + \left[ 1 - 
 \frac{G'(\bbox{x},\tau)}{G'_{max}(\bbox{x})} \right]
 \sum_a \kappa(\bbox{x}) \rho_a(\bbox{x},t) \, .
\label{M1}
\end{eqnarray}
Finally, the orientation relation becomes
\begin{equation}
 \bbox{e}_a(\bbox{x},\tau) = 
\frac{\bbox{\nabla} [ U_a(\bbox{x})
 + V'_{tr}(\bbox{x},\tau) ]} 
 {\| \bbox{\nabla} [ U_a(\bbox{x})
 + V'_{tr}(\bbox{x},\tau) ] \|}
\label{M2}
\end{equation}
with
\begin{equation}
 U_a(\bbox{x}) = - \| \bbox{d}_a/\sigma - \bbox{x} \| \, .
\label{M3}
\end{equation}
Therefore, the average velocity is given by 
\begin{eqnarray}
 \bbox{V}_a(\bbox{x},\tau) &\approx & \frac{1}{\rho_a(\bbox{x},\tau)} 
 \!\!\sum_{\alpha \in a(\tau)}\!\!
 \frac{v_\alpha^0 T(\sigma\bbox{x})} {\sigma} \bbox{e}_a(\bbox{x},\tau) 
 \delta\big(\bbox{x} - \bbox{x}_\alpha(\tau)\big) \nonumber \\
 &\approx & \frac{V^0 T(\sigma\bbox{x})}{\sigma} \bbox{e}_a(\bbox{x},\tau) 
 = \lambda(\bbox{x}) \bbox{e}_a(\bbox{x},\tau) \, ,
\label{M4}
\end{eqnarray}
where $V^0$ is again the average desired pedestrian velocity.  In the
case of {\em frequent} interactions (avoidance maneuvers) of pedestrians,
$V^0$ must be replaced by suitable monotonically decreasing functions
$V_a(\{\rho_c\})$ of the densities $\rho_c$ \cite{Hel2,Grein,Fluid}. Moreover,
fluctuation effects will be stronger, leading to greater diffusion functions
$D_{ab}(\{\rho_c \})$ and broader trails.
\par
Summarizing our results, we have found a macroscopic formulation of trail
formation which is given by equations (\ref{cont}) through (\ref{M4})
with (\ref{V_tr}). Apart from possible analytical investigations, it
allows us to determine implicit equations for the stationary solution, if
the rates $R_a^+(\bbox{x},\tau)$ and $R_a^-(\bbox{x},\tau)$ are
time-independent. Setting the temporal
derivatives to zero, we find the relations
\begin{equation}
 G'(\bbox{x}) = \frac{G'_0(\bbox{x}) + \sum_a \kappa(\bbox{x}) \rho_a(\bbox{x})}
 {1 + \sum_a \kappa(\bbox{x}) \rho_a(\bbox{x})/G'_{max}(\bbox{x})}
\label{former} 
\end{equation}
and 
\begin{equation}
 \bbox{\nabla} \cdot \big[ \rho_a(\bbox{x}) \bbox{V}_a(\bbox{x}) \big]
 = R_a^+(\bbox{x}) - R_a^-(\bbox{x}) \, .
\label{latter}
\end{equation}
Together with equations (\ref{V_tr}) and (\ref{M2}) through (\ref{M4}),
relations (\ref{former}) and (\ref{latter}) allow to calculate the
finally evolving trail system. Again, we see that the resulting state
depends on the two parameters $\lambda$ and $\kappa$. In addition, it is
determined by the respective boundary conditions, i.e. the configuration
and frequency of usage of the entry point-destination pairs, which are
characterized by the concrete form of the entering rates $R_a^+(\bbox{x})$
and leaving rates $R_a^-(\bbox{x})$.
\par 
The advantage of applying the macroscopic equations is that the finally
evolving trail system can be calculated much more efficiently, since
considerably less time is required for computing: The numerical solution
can now be obtained by means of a simple iterative method which is
comparable to the self-consistent field technique.  Examples are shown in
Figure~6 for different values of $\kappa$. As expected, the
results agree with the ones of the related microsimulations, which are
depicted in Figure~5. 

\section{Summary and Outlook}

We showed that the active walker concept is suitable for modeling and understanding
trail formation by pedestrians and animals. Our model 
turned out to be in good agreement with observations. It included
an equation of motion of the walkers, 
an equation describing environmental changes by
the markings which they leave and their decay, 
a relation reflecting the attractiveness
of already existing trails, and an equation delineating their influence on
orientation. Whereas frequently used trails are reinforced, rarely
chosen trails are vanishing in the course of time. This causes a tendency
of trail bundling, which can be interpreted as an agglomeration phenomenon.
However, the evolving
patterns are not localized since the active walkers intend to reach
certain destinations, starting from their respective entry points.
\par
The structure of the resulting trail system can considerably vary with the
species. This depends decisively on the main effect which counteracts the
trail attraction. Whereas our model ants find their destinations (the food sources)
by chance, pedestrians can directly orient towards their destinations,
so that fluctuations are no necessary model component in this case.
Thus, for certain ant species a dendritic trail system is found, the
detailed form of which depends on random events, i.e.
the concrete history of its evolution. 
Pedestrians, however, produce a minimal detour system, i.e. an 
optimal compromise between a direct way system and a minimal way system.
\par
As a consequence, we could derive a macroscopic model for the
trail formation by pedestrians, but not for ants. 
It implied a self-consistent field method for a
very efficient calculation of the finally evolving trail system.
This is determined by the location of the entry points and destinations 
(e.g. houses, shops, or parking lots) and the rates of choosing the possible 
connections between them. Apart from this it depends on two parameters only, 
which was demonstrated by scaling to dimensionless equations. 
These are related to the trail attractiveness and the average velocity of 
motion. 

\subsection{Trail formation as a self-organization phenomenon}

In order to demonstrate that the evolution of trail systems can be
understood as a typical self-organization phenomenon, our model
has made a number of simplifying assumptions about the agents. 
\par
In the example of trail formation by certain ant species, the major
difference to biology is, that the active walkers
used in the simulations have far less complex capabilities than the
biological creatures. They almost behave like physical particles which
respond to local forces in a quite simple manner, without ``implicit and
explicit intelligence'' \cite{haefner-crist-94}.  Compared to the complex
`individual-based' models in ecology \cite{deanged-92}, the active walker
model proposed here provides a very simple but efficient tool to simulate
a specific structure only with a few adjustable parameters.
\par 
With respect to the formation of trunk trails, our model indicates that
these patterns can be obtained also under the restrictions, that (i) no
visual navigation and internal storage of information is provided, (ii)
in the beginning, no chemical signposts exist which lead the ants to
the food sources and afterwards back to the nest. Rather, the formation
of trail systems can be described as a process of self-organization.
Based on the interactions of the active walkers on a local or
`microscopic' level, the emergence of a global or `macroscopic' structure
occurs. The basic interaction between the
active walkers can be considered as indirect communication mediated by an
external storage medium \cite{Schw,Futur}. This is a collective process
in which all active walkers are involved. The information which an active
walker produces in terms of chemical markings affects the behaviors of
the others. It can be amplified during the evolution process or
disappear again, thus leading to a correlation between the information
generated and to the self-organization of the walkers on a spatial level.

\subsection{Implications for urban planners: Optimization of way systems}

Computer simulations of our pedestrian trail formation model will be
a valuable tool for designing convenient way systems (cf. Fig.~7). 
For planning purposes the model parameters $\lambda$ and $\kappa$ must be 
specified in a realistic way. Then, one needs to simulate
the exptected flows of pedestrians that enter the considered
system at certain entry points with the intention to reach certain
destinations. Already existing ways can be taken into account by
the function $G'_0(\bbox{x})$. According to our model, 
a trail system will evolve which minimizes overall detours and thereby
provides an optimal compromise between a direct and a minimal way system.
It is expected that the corresponding ways meet the pedestrian requirements
best: They will most likely be accepted and actually used, since they take 
into account the route choice habits of pedestrians.
For the simulation of realistic situations, the results
can serve as planning guidelines for architects, landscape gardeners, and urban
planners.

\subsection{Current research directions}

Besides of possible applications, our present research focusses on two 
questions: (i) How must our trail formation model 
be specified in order to be applicable to 
trail formation by hoofed animals or mice \cite{Eda,Otto}?
(ii) Can our model be generalized in a way that allows to
understand {\em human decision making}, in particular processes of finding
suitable compromises? Interestingly enough, one
says that someone ``follows in somebody's footsteps'' or that someone
``treads new paths''. Therefore, a related theory 
for a more abstract space (which represents the 
set of behavioral alternatives) may describe the
evolution of social norms and conventions \cite{Hel3}.

\section*{Acknowledgments}

The authors would like to thank K. Humpert 
and B. H\"olldobler for providing some of their interesting 
(photo)graphical material. 
\clearpage
\begin{figure}[htbp]
\begin{center}
\unitlength1cm
\begin{picture}(16,10)
\put(6.5,0){\epsfig{height=10cm, file=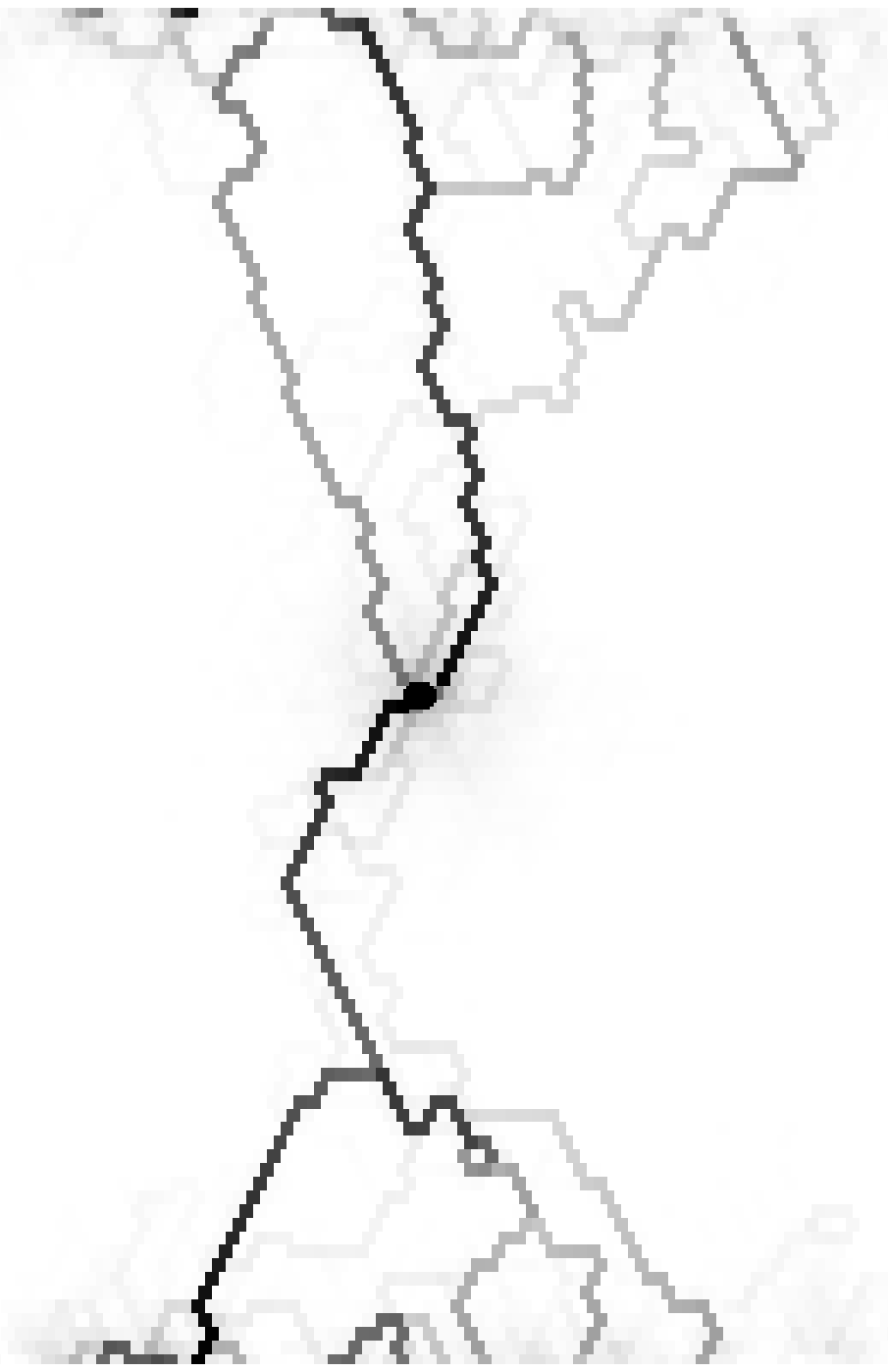}}
\put(-0.5,0){\epsfig{height=10cm,
      bbllx=183pt, bblly=302pt, bburx=413pt, bbury=548pt, 
      file=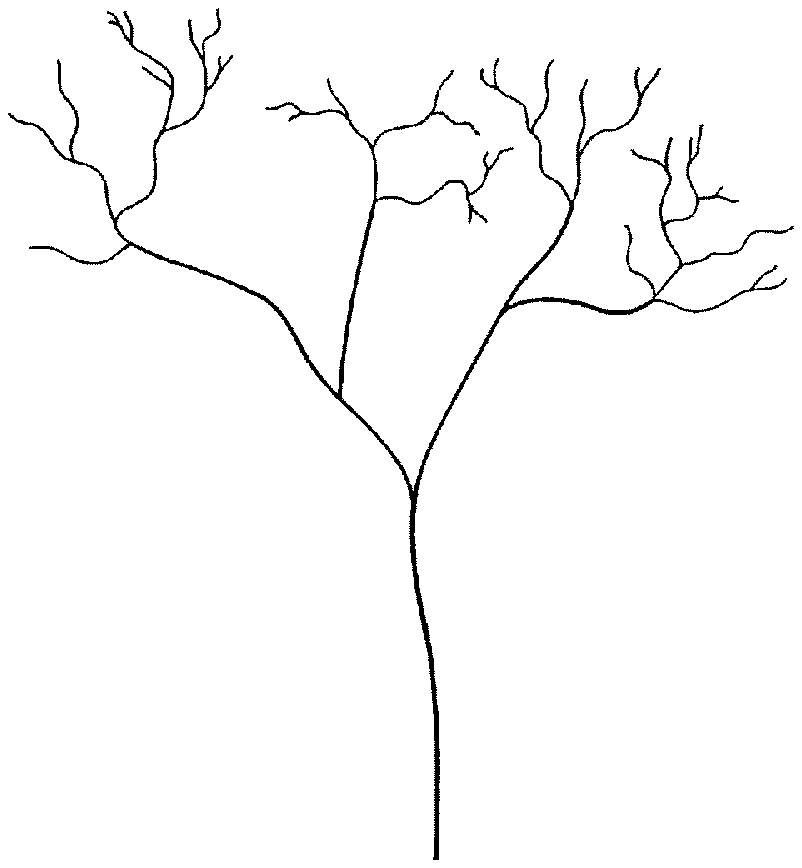}}
\put(7.3,1){\Large (b)}
\put(0,1){\Large (a)}
\end{picture}
\end{center}
\caption[]{(a) Dendritic trunk trail system of the ant species {\it Pheidole
militicida} (after \cite{Hoell}).
(b) Simulation result of trunk trail formation by active walkers. The result is
in good agreement with the empirical findings.}
\label{ants}
\end{figure}
\clearpage
\begin{figure}[htbp]
\begin{center}
\leavevmode
\epsfig{width=10cm, file=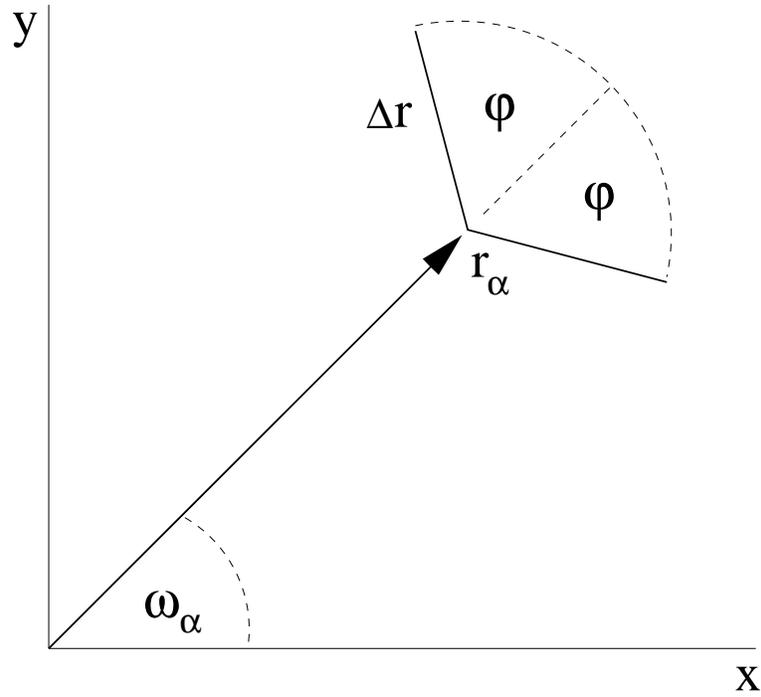}
\end{center}
\caption[]{Illustration of the model quantities characterizing the ant-like active
walker $\alpha$. The arrow represents the body of the ant and its orientation
$\omega_\alpha$ with respect to the axes $x$ and $y$ of the co-ordinate system. 
It ends at the point $\bbox{r}_\alpha$ which corresponds
to the front of the ant's head, where the antennae start. These have the length
$\Delta r$ and include an angle $2\varphi$ of perception.}
\label{ameise1}
\end{figure}
\clearpage
\begin{figure}[htbp]
\begin{center}
    \leavevmode \epsfig{width=15cm, angle=0, 
      bbllx=0pt, bblly=0pt, bburx=131pt, bbury=88pt, 
      file=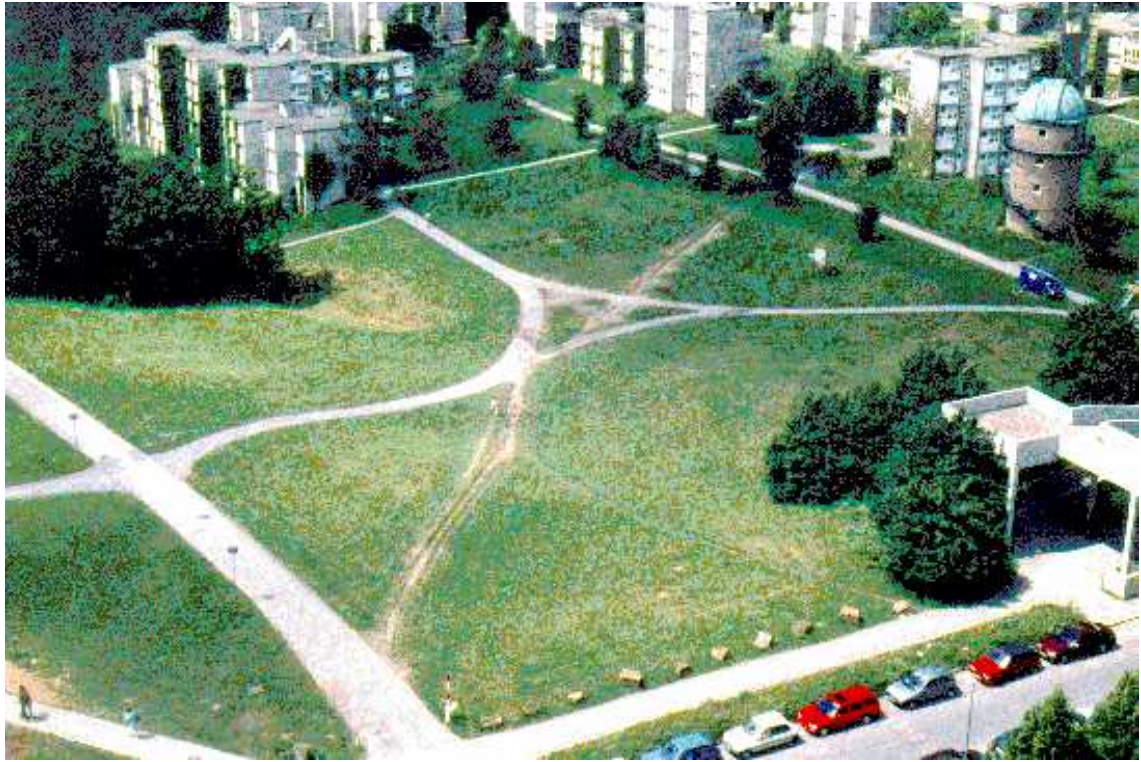}
\end{center}
\caption[]{Between the straight, paved ways on the university campus 
in Stuttgart-Vaihingen a trail system has evolved (center of the 
picture). Two types of nodes are observed: 
 Intersections of two trails running in a straight line and
 junctions of two trails which smoothly merge into one trail
\cite{Nature,Hel2}.}
\label{trail_vaih}
\end{figure}
\clearpage
\begin{figure}[htbp]
\begin{center}
\unitlength1.2cm
\begin{picture}(12.5,11)
\put(0,10.8){\epsfig{height=10.75\unitlength, 
      bbllx=177pt, bblly=350pt, bburx=320pt, bbury=720pt, angle=180, clip=,
      file=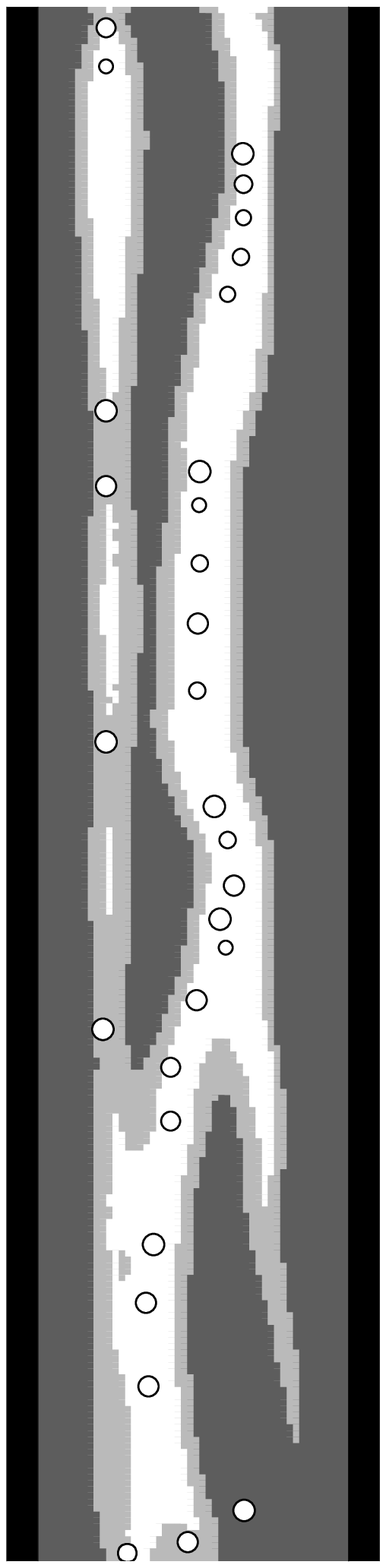}}
\put(4.8,0){\epsfig{height=11\unitlength, angle=0, 
      bbllx=508pt, bblly=0pt, bburx=596pt, bbury=131pt, 
      file=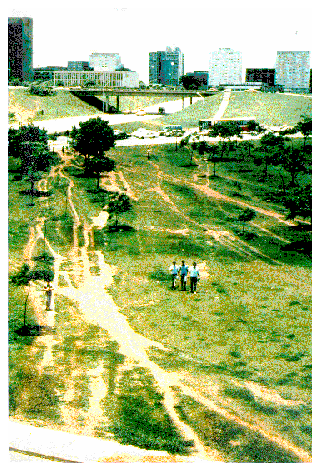}}
\end{picture}
\end{center}
 \caption[]{When pedestrians leave footprints on the ground, trails will develop,
 and only parts of the ground are used for walking (in contrast to
 paved areas). The similarity between the simulation result (left)
 and the trail system on the university
 campus of Brasilia (right, reproduction by kind permission of
 Klaus Humpert) is obvious \cite{Nature,Hel2}.}
\label{tramp}
\end{figure}
\clearpage
\begin{figure}[htbp]
\begin{center}
  \epsfig{width=5.2cm, angle=0, bbllx=274pt, bblly=402pt, bburx=321pt, 
  bbury=449pt, file=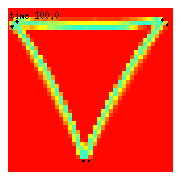}
  \hfill
  \epsfig{width=5.2cm, angle=0, bbllx=274pt, bblly=402pt, bburx=321pt, 
  bbury=449pt, file=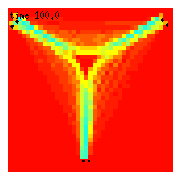}
  \hfill
  \epsfig{width=5.2cm, angle=0, bbllx=274pt, bblly=402pt, bburx=321pt, 
  bbury=449pt, file=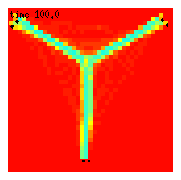}
\end{center}
\caption[]{The structure of the emerging trail system (light grey) 
essentially depends on the attractiveness of the
trails (i.e.\ on the parameter $\kappa = I T/\sigma$). 
If trail attractiveness is small, a direct way system develops
(left), if it is large, a minimal way system is formed,
otherwise a minimal detour system will result (middle) which looks
similar to the trail system in the center of Fig.~\ref{trail_vaih}.
The grey scale allows to reconstruct the temporal evolution of the
trail system before its final state was reached \cite{Hel2}.}
\label{wegtyp}
\end{figure}
\begin{figure}[htbp]
\begin{center}
\unitlength1cm
{\epsfig{width=5.2cm, angle=180, file=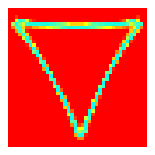}}
  \hfill
{\epsfig{width=5.2cm, angle=180, file=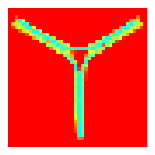}}
  \hfill
{\epsfig{width=5.2cm, angle=180, file=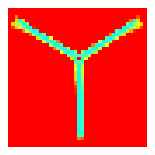}}
\end{center}
\caption[]{Stationary solution
of the macroscopic trail formation model, obtained by an iterative
self-consistent field method. Since the 
boundary conditions were chosen 
as in Figure~\ref{wegtyp},
the results in dependence of the
parameter $\kappa$ are almost identical with those
of the corresponding microsimulations \cite{Nature}.}
\label{mac}
\end{figure}
\clearpage
\begin{figure}[htbp]
\begin{center}
\epsfig{width=12cm, file=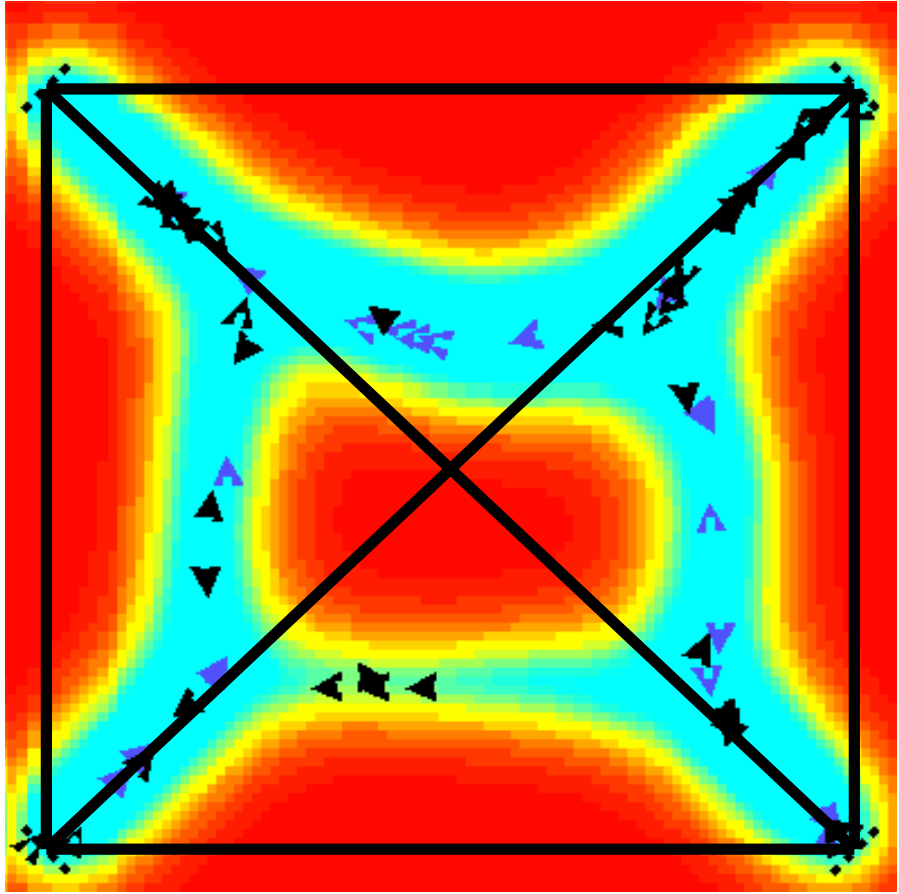}
\end{center}
\caption[]{Comparison of different types of way systems between four places: 
The {\em direct way system} (which is represented by the black lines) provides 
the shortest connections between all entry points and destinations, but
it covers much space. In real situations, pedestrians will
produce a {\em `minimal detour system'} as the best compromise between a direct way
system and a {\em minimal way system} (which is the shortest way system that
connects all entry points and destinations) \cite{Hel2}.
The illustration shows a simulation result 
which could serve as a planning guideline. Its asymmetry is caused by
differences in the frequency of trail usage. (Note that the above figure, in
contrast to Figs.~\ref{wegtyp} and \ref{mac}, 
does not display the ground potential, but
the trail potential. The latter appears considerably broader, since it
takes into account the range of visibility of the trails. Arrows represent the
positions and walking directions of pedestrians. Therefore, they indicate 
the actually taken ways.)}
\label{wegsys}
\end{figure}
\end{document}